\documentclass[journal=apchd5,manuscript=article]{achemso}

\usepackage{chemformula} 
\usepackage[T1]{fontenc} 
\usepackage{graphicx}
\usepackage{dcolumn}
\usepackage{natbib}
\usepackage{amsmath,amsfonts,amssymb,bm}

\author{Huatian Hu}

\altaffiliation{Current address: Istituto Italiano di Tecnologia, Center for Biomolecular Nanotechnologies, Via Barsanti 14, 73010 Arnesano, Italy}
\affiliation[East China Normal University]{State Key Laboratory of Precision Spectroscopy Science and Technology, East China Normal University, Shanghai 200241, China.}
\alsoaffiliation[Wuhan Institute of Technology]
{Hubei Key Laboratory of Optical Information and Pattern Recognition, Wuhan Institute of Technology, Wuhan 430205, China.}
\author{Zhiwei Hu}
\affiliation[East China Normal University]{State Key Laboratory of Precision Spectroscopy Science and Technology, East China Normal University, Shanghai 200241, China.}
\author{Christophe Galland}
\email{chris.galland@epfl.ch}
\affiliation[Ecole Polytechnique Fédérale de Lausanne]
{Institute of Physics, Ecole Polytechnique Fédérale de Lausanne (EPFL), CH-1015 Lausanne, Switzerland.}
\author{Wen Chen}
\email{wchen@lps.ecnu.edu.cn}
\affiliation[East China Normal University]{State Key Laboratory of Precision Spectroscopy Science and Technology, East China Normal University, Shanghai 200241, China.}

\title[An \textsf{achemso} demo]
  {Plasmonic Nanoparticle-on-nanoslit Antenna as Independently Tunable Dual-Resonant Systems for Efficient Frequency Upconversion}

\abbreviations{IR,NMR,UV}

\begin{document}

\begin{tocentry}

Some journals require a graphical entry for the Table of Contents.
This should be laid out ``print ready'' so that the sizing of the
text is correct.

Inside the \texttt{tocentry} environment, the font used is Helvetica
8\,pt, as required by \emph{Journal of the American Chemical
Society}.

The surrounding frame is 9\,cm by 3.5\,cm, which is the maximum
permitted for  \emph{Journal of the American Chemical Society}
graphical table of content entries. The box will not resize if the
content is too big: instead it will overflow the edge of the box.

This box and the associated title will always be printed on a
separate page at the end of the document.

\end{tocentry}

\begin{abstract}

Dual-band plasmonic nanoantennas, exhibiting two widely separated user-defined resonances, are essential for studying and optimizing plasmon-enhanced optical phenomena, including photoluminescence, Raman scattering, and nonlinear effects such as harmonic and sum-frequency generation. The nanoparticle-on-slit (NPoS) or nanoparticle-in-groove (NPiG) antenna is a recently introduced dual-band structure with independently tunable resonances at mid-infrared and visible wavelengths. It has been used to enhance sum- and difference-frequency generation from optimally located molecules by an estimated $10^{13}$-fold. However, theoretical understanding of its eigenmodes remains limited, constraining further optimization and broader application. Here, we investigate the quasi-normal modes (QNMs) supported by NPoS structures, analyzing how both near-field (giant photonic density of states) and far-field (radiation pattern) characteristics influence upconversion. We identify tuning strategies to adjust visible and mid-infrared resonances independently while maintaining strong near-field mode overlap, which is key for nonlinear processes. Additionally, mode analysis reveals a previously unexplored resonance offering greater field enhancement and superior spatial mode overlap with the mid-infrared field, potentially improving upconversion efficiency fivefold compared with the existing results. This work helps to rationalize and optimize the enhancement of nonlinear effects across a wide spectral range using a flexible and experimentally attractive nanoplasmonic platform.
\end{abstract}

\section*{Introduction}
The concept of light confinement is pivotal in nanophotonics, since it enables the control of the intrinsic photophysical properties of emitters, including molecules, quantum dots, and two-dimensional materials, by judiciously designing the extrinsic properties of cavities. Notoriously, the spontaneous emission rate can be modified by the Purcell factor, which scales as $\sim Q/V$, where $Q$ is the optical quality factor and $V$ is the mode volume. \cite{akselrodProbingMechanismsLarge2014a, riveraLightMatterInteractions2020, wuNanoscaleLightConfinement2021b} Plasmonic nanocavities and, more recently, picocavities, can localize the field down to the cubic nanometer scale, offering unprecedented opportunities for boosting light-matter interactions at the nanoscale. \cite{benzSinglemoleculeOptomechanicsPicocavities2016a, baumbergExtremeNanophotonicsUltrathin2019, liBoostingLightMatterInteractions2024a,muhlschlegel2005resonant,novotny2012principles}
To maximize the light-matter interaction and the collected optical signal, several parameters must be optimized \cite{roelli2024nanocavities}: (1) the excitation enhancement of the emitter, which depends on the coupling efficiency of the incoming light to the plasmonic resonance, on the mode volume, and on the local field polarization alignment with the emitters' orientation;
(2) the emission enhancement and light collection efficiency, which require a large radiative local density of states (LDoS) and highly directional far-field radiation patterns.
For optical processes whose excitation and emission frequencies are closely aligned, such as surface-enhanced Raman spectroscopy (SERS), it is sufficient to optimize all parameters above for a single plasmonic resonance. \cite{zhangSimultaneousSurfaceenhancedResonant2019, langerPresentFutureSurfaceEnhanced2020}
However, optical processes that involve well-separated frequencies, such as photoluminescence and nonlinear frequency up- and down-conversion, \cite{thyagarajanEnhancedSecondharmonicGeneration2012, akselrodLeveragingNanocavityHarmonics2015, celebranoModeMatchingMultiresonant2015, noorModeMatchingEnhancementSecondHarmonic2020c} 
can introduce another demand, i.e., (3) the dual-resonant enhancement of excitation and emission processes together with a good field overlap between the two resonant modes. 

To date, a wide range of dual-resonant nanostructures have been developed to optimize the enhancement of various optical processes, including high-frequency SERS,\cite{chuDoubleResonancePlasmonSubstrates2010} exciton luminescence in two-dimensional semiconductors,\cite{chuDoubleResonancePlasmonSubstrates2010} second harmonic generation,\cite{thyagarajanEnhancedSecondharmonicGeneration2012, celebranoModeMatchingMultiresonant2015} and surface-enhanced infrared absorption (SEIRA) for multicomponent molecular analysis.\cite{rodrigoResolvingMoleculespecificInformation2018} 
However, most of these structures typically work in the same or adjacent frequency bands, e.g., visible (VIS) or near-infrared (NIR). 
In contrast, many significant optical processes, such as vibrational sum-frequency generation and high-order harmonic generation, can exhibit frequency separations spanning several octaves, making the design, fabrication, and characterization of dual-resonant antennas more challenging. \cite{roelliMolecularPlatformFrequency2020a, chenContinuouswaveFrequencyUpconversion2021a, xomalisDetectingMidinfraredLight2021, chikkaraddySinglemoleculeMidinfraredSpectroscopy2023a} 
Recently, the interplay of Raman scattering (at VIS/NIR frequencies) and infrared absorption (in the mid-infrared, MIR) has been achieved through the plasmonic antennas, which enable simultaneous studies of SERS and SEIRA, and more specific molecular sensing applications. \cite{leMetallicNanoparticleArrays2008, dandreaOpticalNanoantennasMultiband2013, muellerSurfaceEnhancedRamanScattering2021, sanchez-alvaradoCombinedSurfaceEnhancedRaman2023, oksenbergComplementarySurfaceEnhancedRaman2023a}
Chen \textit{et al.} developed a dual-resonant plasmonic nanogap antenna with a nanoparticle-on-nanoslit (NPoS) geometry that hosts hundreds of molecules within the nanogap, successfully demonstrating continuous-wave coherent frequency conversion from the MIR (9.3 um) to the visible domain based on three-wave mixing. \cite{chenContinuouswaveFrequencyUpconversion2021a} Recently, Redolat \textit{et al.} integrated such a NPoS into dielectric waveguides for on-chip SERS spectroscopy. \cite{redolat2025polarization}.
Compared to the simultaneous implementation of SERS and SEIRA, coherent frequency conversion requires the additional condition of good vector field overlap between the MIR and VIS modes mediated through the $\chi^{(2)}$ tensor, making the design of suitable nanocavities more challenging. 

In this article, we thoroughly analyze the optical properties of NPoS plasmonic multiple-resonance antennas, starting from a quasi-normal mode (QNM) analysis of the modes residing at both VIS to MIR wavelengths.
We find that the NPoS resonances in the VIS to NIR range can be attributed to the mode hybridization between two nanoparticle-on-mirror (NPoM) antennas formed by the same nanoparticle with two side walls of the slit. 
This finding offers a theoretical foundation for tuning the resonance with the same methodology as for other NPoM constructs.
On the other hand, the MIR NPoS resonance depends mostly on the length of the slit. 
These two features guarantee freedom to independently modulate the VIS (or NIR) and MIR resonances: the MIR resonance can be adjusted by the length of the nanoslit, while the NIR resonance can be tuned by the gap size, facet size of the nanoparticle and width of the NPoS. 
Additionally, we analyze the mode overlap between QNMs in the VIS/NIR and that in the MIR. Our results suggest that conversion efficiencies significantly exceeding those experimentally achieved in Ref.~\citenum{chenContinuouswaveFrequencyUpconversion2021a} are possible for an optimal choice of cavity parameters and operation wavelength.
Our work contributes to developing and optimizing next-generation plasmonic nanostructures with multiple resonances from the visible to far-infrared spectrum, providing insights for multi-band photonic applications and devices.

\section*{Results and discussion}
\subsection*{Dual-Resonant NPoS Antenna}
\begin{figure}
  \centering
  \includegraphics[width=13cm]{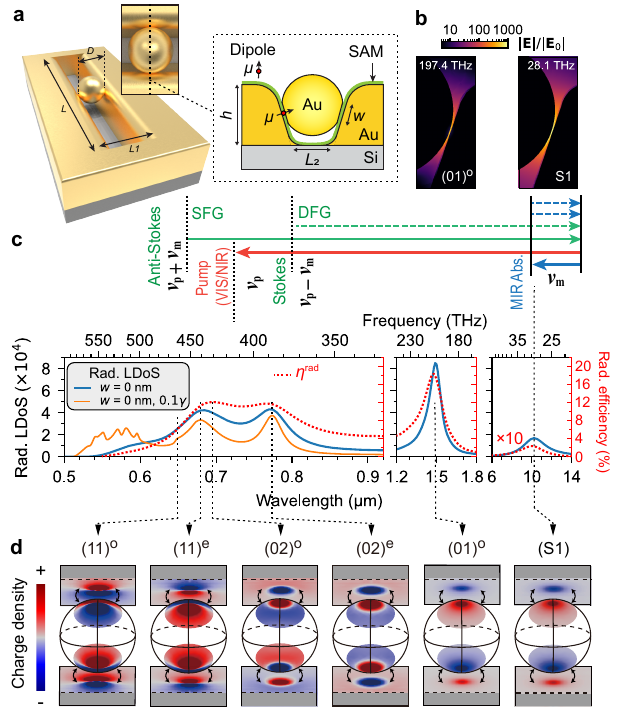}
  
  \caption{\textbf{Quasinormal mode (QNM) analysis and local density of states (LDoS) of the spherical nanoparticle-on-nanoslit (NPoS) systems} (a) Schematic of the structure and definition of its parameters. Geometric parameters: $D=150$~nm, $L=1.4~\mu$m, $L_1=180$~nm, $L_2=80$~nm, {the nanoslit's top and bottom} corners rounded with 30~nm radius, $h=150$~nm, and $w=0$ (spherical NP). (b) Field enhancement over a cross-section of one nanogap under VIS/NIR and MIR plane wave excitation at normal incidence (polarization perpendicular to the slit long axis). (c) Upper panel: schematic of frequency upconversion. Lower panel: radiative LDoS enhancement and corresponding radiative efficiencies ($\eta^{\rm rad}$) of the resonances for a dipole in the center of one of the nanogaps (schematic in (a)). (d) QNM analysis revealing several lower-energy resonances, with corresponding charge distributions. For better visualization, the charges on the sidewalls are moved away from the NP and tilted toward the observer. The nomenclature $(mn)^{\rm p}$ depends on the symmetries and parity of the QNMs. }
  \label{figure1}
\end{figure}

The original design of the dual-resonant antenna was motivated by coherent frequency conversion \cite{chenContinuouswaveFrequencyUpconversion2021a}, which can be described in the general nonlinear optics framework \cite{han2021microwave}. When the nonlinearity is mediated by a molecular vibration, it is also possible to use the formalism of molecular cavity optomechanics \cite{roelliMolecularPlatformFrequency2020a,roelli2024nanocavities}. 
The NPoS structure is comprised of a metallic nanoparticle (diameter of the order of 100~nm) inside a few-micrometer-long slit, whose width matches the nanoparticle diameter and whose sidewalls feature a funnel slope (Fig.~\ref{figure1}a). A {self-assembled monolayer (SAM) of molecules} (or any other thin, low-dimensional material) is acting as a spacer between the nanoparticle and the slit sidewalls. Conceptually, the NPoS realizes a smart integration between a SEIRA-compatible MIR inverted antenna (inspired by the Babinet-principle) \cite{huckPlasmonicEnhancementInfrared2015} and a double-gap NPoM structure with VIS/NIR resonances and extreme field enhancement \cite{baumbergExtremeNanophotonicsUltrathin2019}. 

As shown in Figs.~\ref{figure1}a-c, the NPoS can be resonantly excited by a pair of MIR (signal wave) and VIS/NIR pump fields. The $\chi^{(2)}$ nonlinear response of the nanogap (including molecular and metallic surface contributions) results in the up-converted MIR signals at the Stokes (difference-frequency generation, DFG) and anti-Stokes (sum-frequency generation, SFG) sidebands, both within the VIS/NIR domain. 
These three-wave mixing processes are jointly enhanced by the nanoslit MIR resonance and the gap modes supported by the plasmonic junctions formed between the nanoparticle and the nanoslit sidewalls, akin to the NPoM constructs (Fig.~\ref{figure1}b).
Importantly, the NPoS geometry also achieves the simultaneous confinement of both MIR and VIS/NIR fields in the same nanogaps (Fig.~\ref{figure1}b), resulting in a good mode overlap. 
By optimizing the nanoantenna's radiative efficiencies $\eta^{\mathrm{rad}}_{\mathrm{MIR/SF}}$ and the mode overlap $\eta_{\mathrm{overlap}}$ across all involved frequencies, one can maximize the upconversion efficiency $\eta_{\mathrm{conv}}$ via the relation,
\begin{equation}
\eta_{\mathrm{conv}}\propto \eta^{\mathrm{rad}}_{\mathrm{MIR}}\eta^{\mathrm{rad}}_{\mathrm{SF}}\eta^2_{\mathrm{overlap}},
\label{eq:conversioneffi}
\end{equation}
where the nonlinear field overlap is defined as
\begin{equation}
\eta_{\rm overlap}^2=\frac{\varepsilon_0^3|\int_{\rm gap} {e}^{\rm MIR}_{\bot }(\bm r) {e}^{\rm VIS}_{\bot }(\bm r) {e}^{\rm SF}_{\bot }(\bm r) dV|^2}{V_{\rm MIR}\,V_{\rm VIS}\,V_{\rm SF}\cdot\max\left(\varepsilon_{\rm MIR}|\bm e^{\rm MIR}|^2\right)\cdot\, \max\left(\varepsilon_{\rm VIS}|\bm e^{\rm VIS}|^2\right)\cdot\,\max\left(\varepsilon_{\rm SF}|\bm e^{\rm SF}|^2 \right)} \label{eq:overlap3}.
\end{equation}
Here, $|\bm e^{\rm VIS/SF/MIR}|$, $V_{\rm VIS/SF/MIR}$,  $\varepsilon_{\rm VIS/SF/MIR}$ are the quasinormal modal profiles, mode volumes, and permittivity, respectively, at each wavelength, with SF designating the MIR+VIS sum-frequency. The derivation of this expression is detailed in the Appendix and follows from some simplifying assumptions; the main one is to approximate the field orientation in the nanogap as being strictly normal to the metal surface and to assume that all molecules have the same orientation so that the effective $\chi^{(2)}$ response can be obtained from a single tensor contraction and factored out. 

The two components of an NPoS nanoantenna, namely, a nanoparticle and a nanoslit, can be separately designed to modulate the VIS/NIR and MIR resonances. On the one hand, the MIR resonance is chiefly determined by the length of the nanoslit, while the strength of the local field is dominated in the vicinity of the nanoparticle {due to the nanogap enhancements}. On the other hand, the VIS/NIR plasmonic modes are more sensitive to the morphology of the nanoparticle, the gap size, and the surrounding dielectric environment, while being insensitive to the slit length due to the energy and momentum mismatch. We will substantiate these statements in the following by analyzing the QNMs and calculating the LDoS.

\textbf{Spherical nanoparticle without facets in a slit}. As shown in Fig.~\ref{figure1}a, we first consider a 150~nm-diameter spherical gold  nanoparticle without facets ($w = 0$) mounted on a 1.4~$\mu$m long gold slit covered by a 1~nm thick molecular monolayer, which aligns with the geometry introduced by Chen \textit{et al.} \cite{chenContinuouswaveFrequencyUpconversion2021a} 
The radiative LDoS, which accounts for the radiative channel of the dipole emission by considering nanoantenna's radiative efficiency and the setup collection efficiency $\eta_{\rm coll}$ (see Methods), was calculated in both VIS/NIR and MIR bands (Fig.~\ref{figure1}c, {see the full spectrum in SI Fig. S1}). 
These enhanced vacuum fields are central to boosting light-matter interactions. 
Figure~\ref{figure1}c shows that the NPoS has a MIR resonance near 10~$\mu$m (30~THz) with more than 16,000-fold radiative LDoS enhancement. 
Resonances in the VIS and NIR bands are much more complicated, with three prominent peaks and several shoulders, suggesting the existence of various near-degenerate resonances. 
The radiative LDoS enhancements in the VIS ($\sim 700$~nm) and NIR ($\sim 1500$~nm) are about 40,000 and 80,000, respectively. 
For a clearer observation of the shoulders, we then used in the calculation an artificially small nonradiative decay rate, $0.1\gamma$, shown as the yellow curve of Fig.~\ref{figure1}d. The origin of these shoulders will be clarified by QNM analysis later.

The nanoantenna's radiative efficiency enters the external frequency conversion efficiency. By comparing the radiative LDoS to the total LDoS, we can calculate the radiative efficiency $\eta^{\rm rad}$, which represents the brightness of a resonance. Notably, the resonances in the VIS range typically have an efficiency of around 10\%, while the lower-order mode in the NIR has a slightly higher efficiency of nearly 20\%. The MIR mode is considerably dark with a radiative efficiency of roughly 0.2\% (artificially magnified by 10 times for visualization). 

We perform a rigorous QNM analysis \cite{yan2018rigorous,wu2023modal} (see Methods) to assign the charge distribution (Fig.~\ref{figure1}d) and eigenfrequencies to specific eigenmodes (Table.\ref{tbl:example}). 
Near-degenerate modes typically merged under the broad asymmetric lineshapes are thereby revealed.
{The nomenclature $(mn)^{\rm e/o}$ of each QNM is determined by their symmetries, where the two labels, azimuthal index \textit{m} and radial \textit{n}, correspond to the $m$ pair(s) of azimuthal node(s), and $n-1$ radial node(s), respectively. The mode profiles are analysed in a plane contained in the nanogap between the nanoparticle's facet and the slit's sidewall, and the determination of the ($mn$) follows the same rules as in conventional NPoM systems \cite{tserkezisHybridizationPlasmonicAntenna2015} and is similar to cylindrical optical fibers. (See NPoS's field profile in SI Fig. S2.)}
The superscript labels the mode parity against a vertical plane containing the long axis of the slit: "e" for even ("o" for odd) designates the symmetry (anti-symmetry) of the charge distributions around the two nanogaps. 
In Fig.~\ref{figure1}d, two even modes, $(11)^\mathrm{e}$ and $(02)^\mathrm{e}$, and three odd modes, $(01)^\mathrm{o}$, $(11)^\mathrm{o}$, and $(02)^\mathrm{o}$ are plotted, which dominate the radiative LDOS enhancement in the VIS/NIR. {The (01)$^{\rm e}$ mode is also present, although it is much weaker and typically appears deeper towards the MIR range, around 2.3~$\mu$m as shown in Fig.~S1. For this reason, we do not analyze or discuss this mode in detail in this work.} In addition, the charge distribution of the MIR nanoslit mode (S1) is presented for comparison. As previously discussed, the presence of the nanoparticle concentrates the electromagnetic field within the nanogap, enabling substantial overlap with the VIS/NIR mode. {However, because the nanoparticle does not resonate at the MIR wavelength due to its nanoscale dimensions, this QNM is not governed by the nanoparticle. Instead, it should be interpreted as a perturbed Babinet-type mode dominated by the nanoslit. Although it exhibits a $(01)^{\rm o}$-like symmetry and can be categorized within the same classification framework (as $(01)^{\rm o/slit}$), it should be noted that this $(01)^{\rm o}$ mode is excited in a highly detuned manner. The field profile of the whole NPoS structure is presented in SI Fig. S3, which shows how a nanoparticle's dipole perturbs the nanoslit's MIR mode and its comparison with the nanocavity-determined VIS/NIR modes, e.g., $(01)^o$.}

\begin{table}
  \caption{Quasinormal mode analysis for the NPoS}
  \label{tbl:example}
  \begin{tabular}{cccc}
    \hline
    NPoS QNMs   & Eigenfrequency/THz & Wavelength/nm &$Q$\\
    \hline
     MIR Slit (S1)    & 28.139+4.181i & 10424-1548.6i & 3.4\\
    (01)$^{\rm o}$ & 197.41+12.02i& 1513.0-92.10i& 8.2\\
    (02)$^{\rm e}$ & 386.34+13.66i & 775.02-27.4i   & 14.1 \\
    (02)$^{\rm o}$ & 430.94+38.03i & 690.30-60.92i & 5.7 \\
    (11)$^{\rm e}$ & 448.61+11.90i & 667.80-17.71i & 18.8 \\
    (11)$^{\rm o}$ & 463.88+11.44i & 645.88-15.93i & 20.3 \\
    (03)$^{\rm e}$ & 488.29+26.11i & 612.21-32.74i & 9.3 \\
    (03)$^{\rm o}$ & 516.31+26.02i & 579.17-29.18i & 9.9 \\
    \hline
  \end{tabular}
\end{table}

\begin{figure}
  \centering
  \includegraphics[width=0.9\textwidth]{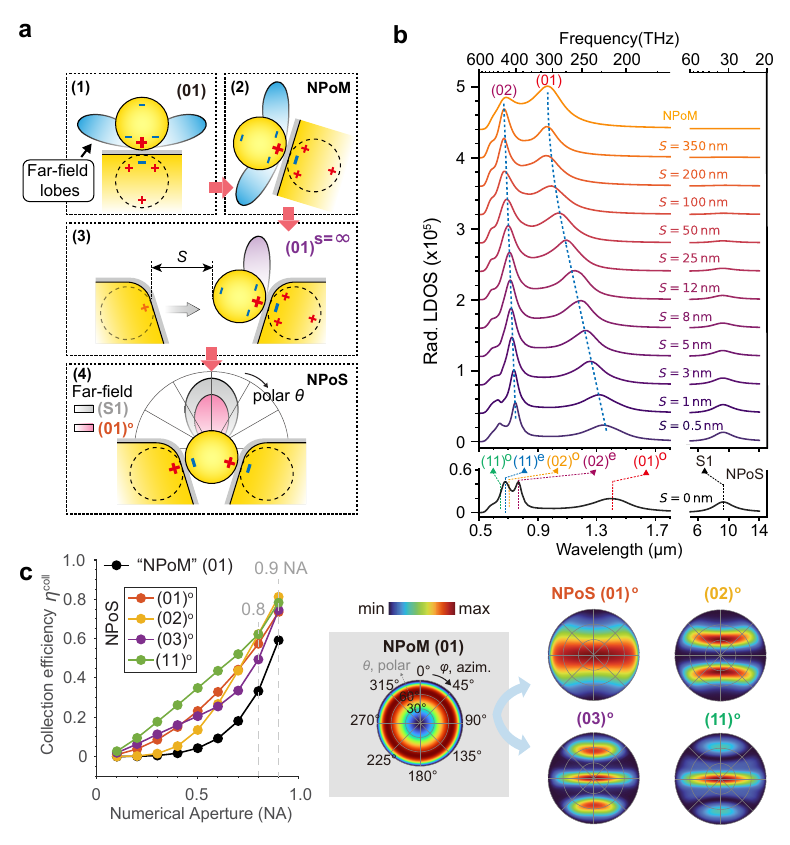}
\caption{\textbf{Plasmonic hybridization in NPoS} (a) Schematics 1 and 2 illustrate the horizontal and tilted NPoM geometries, respectively. Schematics 3 and 4 show the definition of the parameter $S$ and the convergence to the NPoS geometry as $S$ decreases to $\sim1$~nm. Far-field lobes are overlapped to demonstrate the emissive direction.
(b) Radiative LDoS enhancement spectra as a function of $S$, with the horizontal NPoM geometry of the same size and gap thickness plotted as a reference (top yellow curve). The spectra are offset for visualization. 
(c) Far-field radiation patterns, and collection efficiency $\eta_{\rm coll}$ vs. numerical aperture of the horizontal NPoM (01) mode and of several NPoS modes (e.g., (01)\textsuperscript{o}, {$(02)^{\rm o}$}). The azimuthal angle of 0 degree is parallel to the long axis of the slit. All the far-field radiation plots share the same labels and ticks.}
\label{figure2}
\end{figure}

These VIS/NIR QNMs have very similar profiles to those of NPoM constructs (depicted in Fig.~\ref{figure2}a, schematic 1);\cite{tserkezisHybridizationPlasmonicAntenna2015, kongsuwanPlasmonicNanocavityModes2020a} they can be understood as two NPoM junctions sharing the same nanoparticle. 
To prove that, in Fig.~\ref{figure2}, we fix the nanoparticle onto one sidewall of the nanoslit (Fig.~\ref{figure2}a, schematic 2), to form an NPoM-like geometry with the slit sidewall acting as the mirror {(we did not round the slit's bottom edge in this figure to avoid the coupling and anti-crossing with the (01)$\rm ^o$ mode, as discussed in SI Fig. S6)}. Then, as shown by the schematic 3 of Fig.~\ref{figure2}a, we tune the distance $S$ between the other sidewall of the slit and the nanoparticle to control their near-field coupling. 
When $S$ {is sufficiently large (e.g., 350 nm in Fig. \ref{figure2}b)}, the NPoS is {qualitatively} equivalent to a tilted NPoM, with the resonances in the radiative LDoS spectrum very close to those of a horizontal NPoM. {The discrepancy is because the field of nanoparticle can interact with the rounding of the nanoslit, while in the NPoM, the substrate is flat and infinitely large.} 
Going from longer to shorter wavelengths, the 150~nm-diameter horizontal NPoM features two broad resonances plus a shoulder, corresponding to QNMs (01), near-degenerate (11) and (02), and (03), respectively. \cite{tserkezisHybridizationPlasmonicAntenna2015} 
As $S$ decreases, the tilted NPoM couples to the second mirror (opposite sidewall of the slit).
Sharper peaks emerge and redshift in the 500–800~nm range, originating from the coupling of NPoM's (02), (11), and (03) modes to their image charges (as schematic 4 shows in Fig.~\ref{figure2}a). 

For $S<100$~nm, a QNM appears in the NIR and redshifts to 1.4~$\mu$m with $S$ decreasing to 0.5~nm  (increase of the interaction). The spectrum with $S=0$ is plotted at the bottom panel for reference. 
As illustrated in Fig.~\ref{figure1}d, (01)\textsuperscript{o} has a head-to-tail, bonding dipole-dipole interaction,\cite{prodan2003hybridization} which explains the redshift as the interaction increases. 
Compared to the (01) mode in conventional NPoM, (01)\textsuperscript{o}  is unique to NPoS where one nanoparticle couples to two mirrors (slit's sidewalls), and offers a strong radiative resonance at telecommunication wavelengths. 
More importantly, this alignment of the dipoles generates a larger total dipole moment, which correlates with the gradually increasing brightness of the (01)\textsuperscript{o} mode  ($\eta^{\rm rad}\simeq20\%$ revealed in Fig.~\ref{figure1}c), reaching about twice that of the other QNMs at VIS wavelengths.  


The far-field radiation pattern is another important factor in external conversion efficiency since it determines how much energy can be collected. 
The NPoM's (01) mode has a doughnut-shaped far-field radiation diagram around the mirror normal, as depicted in Fig.~\ref{figure2}a), where only a small portion of the energy can be collected by a typical objective lens, whose numerical aperture (NA) is swept until 0.9 in Fig.~\ref{figure2}c. 
In contrast, as illustrated in schematics (2, 3) of Fig.~\ref{figure2}a, the tilted NPoM allows for a larger fraction of the radiation to be collected. 
Similarly, the NPoS geometry ensures a balance of the emission lobes in the far field. The odd modes maintain a dominantly normal emission, with the collection efficiency of the (01)\textsuperscript{o} mode reaching 80\% at NA = 0.9. 
Finally, the S1 mode in the MIR shows a similar emission pattern as the (01)\textsuperscript{o} mode (see Fig.~\ref{figure2}a(4)), due to the same type of charge distribution (see Fig.~\ref{figure1}d), which largely facilitates frequency conversion by using these two modes as dual resonances.

So far, we have revealed a low-order (01)\textsuperscript{o} resonance at telecom wavelengths that has not been experimentally utilized. \cite{chenContinuouswaveFrequencyUpconversion2021a} This bonding dipole-dipole bright mode has a stronger radiative LDoS enhancement than the rest of the higher-order modes, and its far-field diagram shows a convenient normal emission. Since the mid-infrared resonance at $\sim 10~\mu$m depends on the slit length, the change of $S$ (width) has little influence on its frequency but does alter its intensity.
The discovery of the fundamental (01)\textsuperscript{o} resonance motivates future experiments to reach a higher upconversion efficiency, as we will confirm later in Fig.~\ref{figure5} by computing the nonlinear mode overlap.

\begin{figure}
  \centering
  \includegraphics[width=18cm]{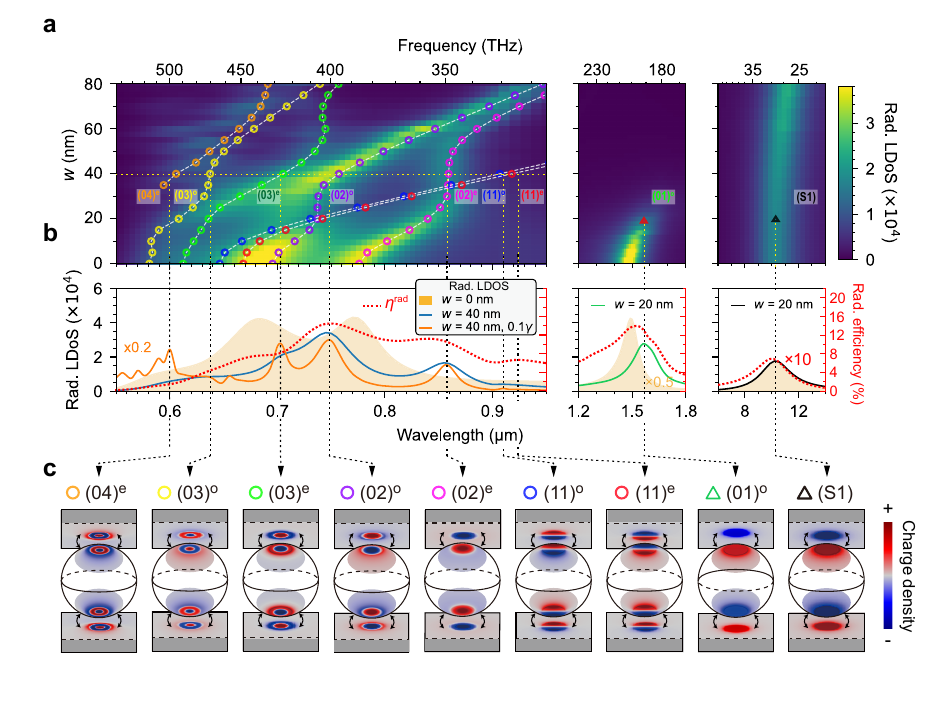}
  \caption{\textbf{NPoS with facets showing hybridized cavity resonances.} (a) Radiative LDoS plotted against facet size on the vertical axis. Calculated QNM eigenfrequencies are marked as open circles, with colors matched to panel (c), and they show excellent correspondence with the peaks in the radiative LDoS. (b) Radiative LDoS spectrum with $\eta^{\rm rad}$ for 40~nm facets (blue line), and the same with reduced broadening (orange line) to better identify different modes. The spherical NPoS ($w=0$) LDoS is plotted as a yellow-shaded area for reference. (c) Induced surface charge densities of each computed QNM.}
  \label{figure3}
\end{figure}

\textbf{Nanoparticle with flat facets.} The nanoparticles studied above were assumed to be perfect spheres, forming longitudinal antenna modes with the substrate\cite{tserkezisHybridizationPlasmonicAntenna2015, kongsuwanPlasmonicNanocavityModes2020a}. 
Synthetic gold or silver nanoparticles often show polyhedral shapes with facets, which can significantly alter the optical response by supporting transverse cavity modes, where most of the energy is confined within an ultrathin nanocavity.\cite{bedingfield2023multi,baumbergExtremeNanophotonicsUltrathin2019} 
In Fig.~\ref{figure3}a, we plot the dependence of the radiative LDoS enhancement spectrum on the facet diameter $w$, increasing from bottom to top.
We assume symmetric facets on both sides of the nanoparticle, parallel to the sidewalls (schematic of the cross-section in Fig.~\ref{figure1}{a}, with $w\neq0$). 
In our approach, nanoparticles with larger facet sizes have a smaller physical volume, thereby falling deeper into the nanoslit, since the molecular spacer thickness is fixed at 1~nm.

As the facet of the nanoparticle grows larger, all modes gradually redshift due to the increased coupling with the mirror and the emergence of \textit{transverse} cavity modes. These \textit{transverse} cavity QNMs have been widely documented in the NPoM-type geometries as the flat bottom facets grow larger to form a metal-insulator-metal cavity with the mirror \cite{tserkezisHybridizationPlasmonicAntenna2015}.
Rich crossing and anti-crossing behaviours are observed in the VIS-NIR domain, where the latter could be explained as the strong coupling between different modes when their near-field patterns have similar symmetry. \cite{tserkezisHybridizationPlasmonicAntenna2015}, such as the QNMs $(mn)^\textrm{o/e}$ with $m=0$. For example, $(02)^o$ and $(03)^e$ near $w=40$ nm, $(03)^o$ coupled with $(03)^e$ and $(04)^e$ around  $w=30$ nm and  $w=55$ nm, respectively. In contrast, the QNMs with different symmetries, such as $m =1$ and $m=0$, can not couple with each other, facilitating crossings in the mapping (e.g., $(11)^\mathrm{o,e}$ both cross $(02)^\mathrm{o,e}$ around  $w=20$ nm and  $w=30$ nm, respectively.
When the facet size $w$ gets significantly larger, the degeneracy of the modes in the VIS range is lifted, and a rich series of cavity modes is observed; e.g, seven prominent modes for $w = 40$~nm (Figs. \ref{figure3}a-c). 

The eigenfrequencies of the QNMs as a function of $w$ (markers with different colours shown in Fig. \ref{figure3}a) are overlapped on the radiative LDoS spectra, showing good alignment with the resonances. Close to anti-crossings, the eigenfrequencies of the hybridized modes can slightly deviate from the radiative LDoS peaks, which has been explained by coupled oscillator models. \cite{pelton2019strong}
The modes are confined in a cylinder-shaped metal-insulator-metal cavity, with mode profiles closely following Bessel functions. 
But the NPoS cannot be simply understood as a slanted NPoM, due to an extra degree of freedom induced by the mode symmetry across the two nanogaps. 
Pairs of modes with opposite symmetries tend to be near degenerate for large facet sizes, but non-degenerate when $w$ is small. 
For instance, (02)\textsuperscript{\textsuperscript{e}} and (02)\textsuperscript{\textsuperscript{o}} have a large frequency difference (>30~THz) when $w = 0$, but they converge for $w \geq 60$~nm. Similarly, 
(11)\textsuperscript{\textsuperscript{e}} and (11)\textsuperscript{\textsuperscript{o}} converge for $w \geq 20$~nm, and that of (02)\textsuperscript{\textsuperscript{e}} and (02)\textsuperscript{\textsuperscript{o}} for $w \geq 70$~nm. 

We can rationalize this behaviour again from the point of view of two coupled NPoMs (Fig. \ref{figure2}b). 
When the $w$ is too small to support transverse modes, all modes should be longitudinal antenna-like. \cite{esteban2015morphology,tserkezisHybridizationPlasmonicAntenna2015, kongsuwanPlasmonicNanocavityModes2020a} 
Therefore, when the tilted NPoM forms an NPoS as schematically shown in Fig.~\ref{figure2}a(3), the evanescent field on the top of the nanoparticle will interact quite strongly with the mirror on the other side. 
On the contrary, when the facet increases, the QNM energy tends to be concentrated in the nanogap cavity with negligible field outside \cite{esteban2015morphology,tserkezisHybridizationPlasmonicAntenna2015, kongsuwanPlasmonicNanocavityModes2020a}, reducing the interaction between {the bare modes (NPoM-like QNMs)}, which become near degenerate.
To sum up, this dependence can be well explained by the mode hybridization theory \cite{prodan2003hybridization} and the evolution of the modes from longitudinal (antenna-like) to transverse (cavity-like) with increasing $w$.

{Last but not least, from the energy of each QNM in Fig.~\ref{figure3}a, we can easily see anti-crossing behaviours between two modes with the same azimuthal index $m$ but with different radial index $n$, e.g., 
(02)\textsuperscript{o} and (03)\textsuperscript{e}, (03)\textsuperscript{o} and (04)\textsuperscript{e}.
But (11)\textsuperscript{o,e} on the other hand, can cross all these modes with $m=0$ due to the near vanishing vector field overlap. Very similar phenomena have been observed in the hybridization of a single NPoM, \cite{tserkezisHybridizationPlasmonicAntenna2015} where the bare modes with different azimuthal order $m$ cross each other while the modes with the same $m$ tend to interact.}
{The radiative LDoS in Fig. \ref{figure3}a (see Methods) was calculated by placing a single dipole on one side of the two nanogaps, thereby breaking the mirror symmetry of the NPoS structure and allowing the excitation of both even and odd parity modes. As a result, the evolution of the QNMs can be clearly visualized in the LDoS maps.} 

{It should be noted that the geometry considered here, a nanoparticle with two facets parallel to the sidewalls, represents a simplified and idealized case. When considering the exact polyhedral shape of nanoparticles and their finer local geometric features (corners, protrusions, etc.), the inherent randomness in their orientations and contact with the nanoslits will significantly impact the NPoM-like QNMs. For example, if the nanoparticle facets are not perfectly parallel to the sidewalls, the contact could become point-like~\cite{huang2016ultrasmall}, effectively increasing the nanogap thickness, decreasing the effective facets, and resulting in a blueshift of both the bare modes and the hybridized plasmons (see SI Fig. S5). Note that this is just one specific case, since the hybridization model for generating $(mn)^{\rm o,e}$ QNMs is general, the nanoparticles with altered geometries still follow the same coupling mechanism, albeit with slightly different oscillator frequencies and coupling strengths. This work aims to propose a general theoretical framework for NPoS's QNMs, and the other crucial factors, such as nanoparticle's exact shape, asymmetric nanogaps, relative orientation and position to the nanoslit, remain an open problem for future exploration.}

\begin{figure}[ht]
  \centering
  \includegraphics[width=16cm]{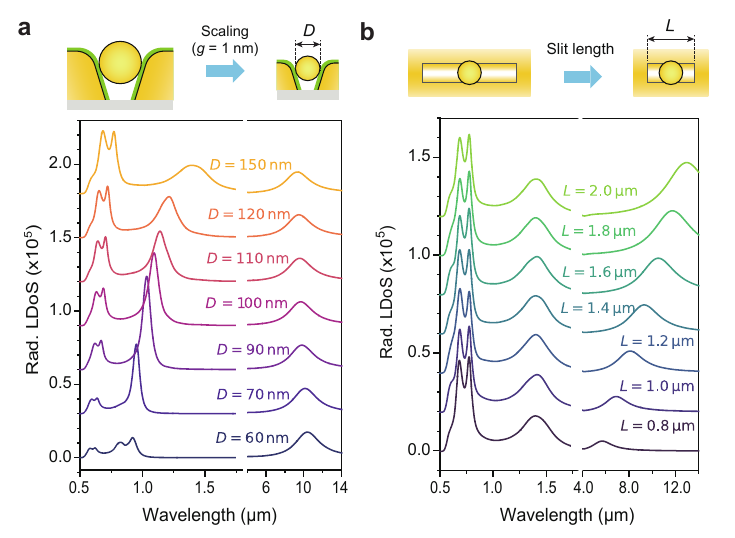}
  \caption{\textbf{Independently tunable NPoS resonances} Dependence of the radiative LDoS spectra of NPoS on (a) the nanoparticle diameter and slit width, and (b) the slit length. In (a), the cross-section of the NPiG is scaled except for the gap thickness that is fixed (1 nm) to hold one layer of molecules. {The length of the slit in (a) is 1.4 $\mu$m.} 
  Offset is 30000 for (a) and 20000 for (b).}
  \label{figure4}
\end{figure}

\textbf{Independent tuning of MIR and VIS/NIR resonances}. 
As shown in Fig. \ref{figure4}a, the size of the particle is a convenient handle to tune the VIS/NIR plasmonic resonances. 
Since the width of the slit must be adapted to hold the nanoparticle, we scale the cross-sectional geometry of the NPoS while keeping the gap thickness at 1 nm (determined by the spacer molecules). The length of the slit is unchanged {at 1.4 $\mu$m} to maintain the MIR resonance at 10 $\mu$m. {As shown by the schematic of Fig. \ref{figure4}, we did not round the bottom corners of the nanoslit in this figure, following the same reason as Fig. \ref{figure2} and Fig. S6, for a clearer demonstration of the plasmon shift.}
As the nanoparticle diameter decreases below 150~nm, both VIS and NIR modes blueshift, while the MIR resonance hardly changes. 
This means that reducing the size (scaling down the cross-sectional geometry) can effectively tune the VIS resonance independently while ensuring the stability of the MIR resonance. 

The tunability of the MIR mode is essential to match the vibrational frequencies of various molecules. 
In Fig. \ref{figure4}b, the MIR resonance is tuned from 6 $\mu$m to 15 $\mu$m as the slit length $L$ changes from 1 $\mu$m to 2 $\mu$m while the VIS/NIR resonances remain unaffected. 
{This can also be explained from the mode profile. We plotted the full electric field profiles of the NPoS while varying the slit length in SI Fig. S4. The MIR mode (S1) consistently exhibits a strong field distributed along the slit, as it originates from the bare slit mode (SI Fig. S3), causing the field profile to expand with increasing slit length. In contrast, the NIR (01)$^{\rm o}$ mode remains tightly confined within the nanogap, with only a weak background field in the slit, especially for longer slits.}
This straightforward independent tuning of VIS/NIR and MIR responses, while maintaining near-field overlap, is a key asset of the NPoS geometry for nonlinear nano-optics and molecular spectroscopy.

\subsection*{Mode Overlap and conversion efficiencies}
\begin{figure}[ht!]
  \centering
  \includegraphics[width=0.6\textwidth]{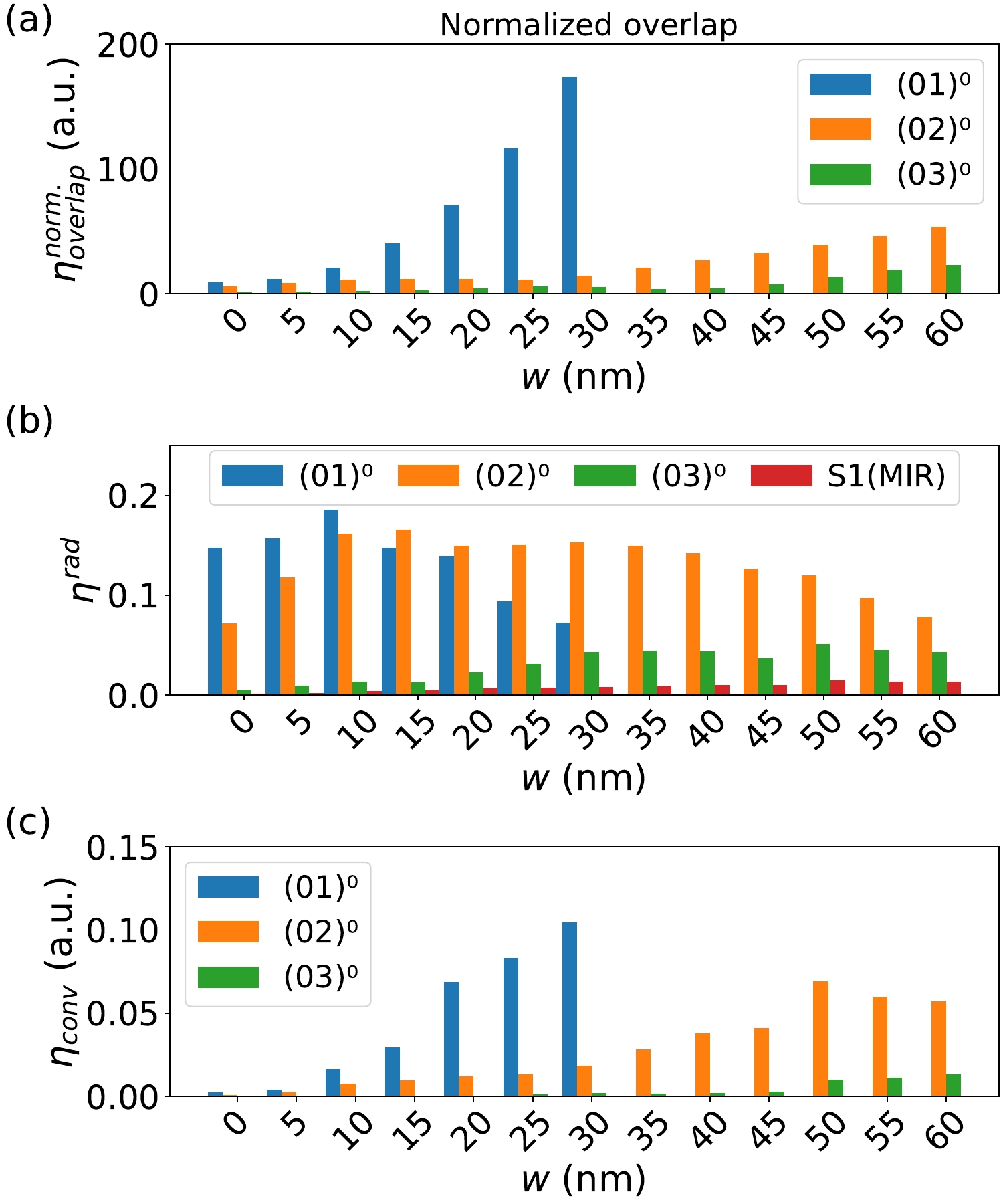}
  \caption{\textbf{
  Nonlinear overlap efficiency of the MIR mode with other NIR and VIS modes.} (a) Normalized mode overlap $\eta^{\rm norm}_{\rm overlap}$ between mode S1 and modes 01$^{\rm o}$, 02$^{\rm o}$, 03$^{\rm o}$, with $w$ varied from 0 to 60 nm. Here, the overlap is normalized to be dimensionless by the wavelengths, $ \eta_{\rm overlap}^{\rm norm.} = \eta_{\rm overlap}^2 \lambda_{\rm SF}^2 \lambda_{\rm MIR}\propto \mathcal{C} $. (b) The corresponding radiative antenna efficiencies $\eta^{\rm rad}$ (c) and relative external frequency upconversion efficiency calculated as $\eta_{\rm conv}\propto\eta^{\rm rad}_{\rm SF} \eta^{\rm rad}_{\rm MIR} \eta_{\rm overlap}^{n\rm orm.}$ }
  \label{figure5}
\end{figure}

Mode overlap is a crucial condition for efficient nonlinear frequency conversion. Figure \ref{figure5}a shows the dimensionless, normalized mode overlap efficiency $ \eta_{\rm overlap}^{\rm norm.} = \eta_{\rm overlap}^2 \lambda_{\rm SF}^2 \lambda_{\rm MIR} $ between each VIS or NIR QNM with the MIR resonance. The definition of $\eta_{\rm overlap}$ is presented in the Appendix, the main simplifying assumptions being that the electric field is locally perpendicular to the metal surface and the nonlinear medium is homogeneously distributed in the nanogaps. Since the MIR mode (S1) is antisymmetric (odd), in principle, only odd VIS/NIR modes (e.g., (01)\textsuperscript{o}, (02)\textsuperscript{o}, (03)\textsuperscript{o}, etc.) can have nonzero mode overlap with S1. However, in the experiment with non-uniform molecule distribution and complex nanoparticle morphology, $\eta_{\rm overlap}$ could be nonzero for even modes as well due to the broken symmetry.

As shown in Fig.~\ref{figure5}a, with increasing facet size, $\eta_{\rm overlap}^{\rm norm.}$ of each QNM with the mid-infrared mode keeps increasing, where the fundamental NIR (01)\textsuperscript{o} mode has the best overlap efficiency. This mode vanishes after $w\geq 30$ nm due to over-damping (see Fig. \ref{figure3}). In general, we find that lower-order modes with longer wavelength (i.e., smaller wavelength mismatch with the MIR) have the larger normalized overlap ($\rm (01)^o>(02)^o>(03)^o$). Enlarging the facet to form a plasmonic nanocavity constitutes the first design rule for optimizing the overlap efficiency. We do not investigate the (11)\textsuperscript{o} here since it is a dark mode (see radiative LDoS in Fig.~\ref{figure3}) and will not have a significant contribution to the upconversion. 

Figure~\ref{figure5}b shows the antenna efficiency $\eta^{\rm rad}$ of the NPoS at the eigenfrequency of each QNM. While each VIS/NIR mode presents an optimal radiative efficiency at a certain $w$, the MIR mode shows a monotonic increase of $\eta^{\rm rad}$ with facet size. The first two lower-order resonances (i.e.,  (01)\textsuperscript{o} and (02)\textsuperscript{o}) show a comparable optimized $\eta^{\rm rad}$ around 10\%-15 \%. {Due to the overdamped nature of the (01)$\rm ^o$ mode for $w> 30$ nm, it is hard to rigorously estimate the antenna efficiency: the modes nearby have non-negligible contributions to the dipole radiation, which violates the single-mode approximation. We chose to stop calculating $\eta^\mathrm{rad}$ and $\eta^\mathrm{conv}$ from there. In future studies, a more rigorous estimation of the conversion efficiency may involve directly deriving the radiative efficiency from the QNM analysis.}

The overall external MIR to VIS/NIR conversion efficiency, plotted for several modes in Fig. \ref{figure5}c according to the formula defined in Eq. \ref{eq:conversioneffi} and thus $\eta_{\rm conv}\propto\eta^{\rm rad}_{\rm SF} \eta^{\rm rad}_{\rm MIR} \eta_{\rm overlap}^{n\rm orm.}$  (see Appendix), incorporates all factors calculated above, reflecting a combination of good near-field mode overlap and far-field radiative efficiency. For small facets, the fundamental NIR (01)\textsuperscript{o} mode harbours the best efficiency, whereas the (02)\textsuperscript{o} mode is optimal for larger facets.
Interestingly, the lowest order resonance (01)\textsuperscript{o} was not addressed in previous experiments \cite{chenContinuouswaveFrequencyUpconversion2021a} as it was out of the investigated frequency band. Our estimation (Fig. \ref{figure5}c) suggests that addressing this mode at telecom wavelength can lead to more than 5 times improvement in conversion efficiency over previous results based on the (02)\textsuperscript{o} mode at shorter wavelengths.


\section*{Conclusions}
We investigated the QNMs of the recently proposed NPoS antenna as a promising nanophotonic structure for efficient optical frequency conversion between mid-infrared and visible domains. We obtained a detailed understanding of the origin of the multiple resonances in the LDoS spectrum and of their behaviours as a function of the NPoS morphology and dimensions. We compute a giant dual-band enhancement in the radiative LDoS and propose a simple yet effective approach to independently tune the MIR and VIS/NIR modes. 
We also predict a previously unreported fundamental mode at telecom wavelengths that offers superior overlap with the MIR mode and enhanced radiative efficiency. This new mode enables significantly higher upconversion efficiency compared to QNMs at shorter wavelengths. Overall, our study rigorously evaluates the NPoS antenna as a powerful and timely platform for nonlinear nanophotonics, while uncovering a new route to pushing upconversion efficiency beyond current limits.

\section*{Methods}
\textbf{Structure.} We simulated the NPoS antenna with a finite-element method package COMSOL Multiphysics. As shown in {Fig.~\ref{figure1}a}, the cross-section of the nanoslit is made of a trapezoidal void in a $h=150 \,\mathrm{nm}$ thick gold film, whose two parallel sides were $L_1=80 \,\mathrm{nm}$ (bottom) and $L_2=180 \,\mathrm{nm}$ (top). The corners of the nanoslit were rounded by a 30 nm radius of curvature to ensure convergence. A 1 nm thick spacer was assumed as the nonlinear material (typically, the molecular layer). This cross-section was extruded into a long slit with length $L_\mathrm{}$. A gold nanoparticle with diameter $D$ was finally mounted into the slit, forming an NPoS antenna. To capture the realistic morphology of a gold nanoparticle, facets of diameter $w$ were introduced. The facets were parallel to the sidewalls of the nanoslit. The whole geometry was covered by a 200 nm thick perfectly matched layer as boundary conditions {(5 layers of mesh).} The permittivity of the spacer was $\varepsilon_{\mathrm{mole}} = 1.4^2$ and that of silicon $\varepsilon_{\mathrm{Si}} = 4^2$. The parameters of the Lorentz-Drude model for gold were $\varepsilon(\omega) = \varepsilon_{\infty}-\varepsilon_{\infty} \sum_i \omega_{p, i}^2\left(\omega^2-\omega_{0, i}^2+i \omega \gamma_i\right)^{-1}$ were: $\varepsilon_{\infty} = 6$; Lorentz term: $\omega_{p, 1}$ = $2\pi\cdot360.3$ THz, $\omega_{0,1}$ = $2\pi\cdot727.6$ THz, $\gamma_{1}$ = $2\pi\cdot212.0$ THz; Drude term: $\omega_{p, 2}$ = $2\pi\cdot854.7$ THz, $\omega_{0,2}$ = 0, $\gamma_{2}$ = $2\pi\cdot9.9$ THz.\cite{wu2023modal}

\textbf{Local density of states.} To compute the LDoS enhancement, we excited the system with an electric point dipole in one of the nanogaps (schematic see Fig.\ref{figure1}a). The dipole moment $\mathbf{\mu}$ is normal to the sidewall. The LDoS enhancement is $\rho/\rho_0 = P/P_0$, where $\rho$ (resp., $\rho_0$) is the LDoS with (resp., without) plasmonic structure, and $P$ (resp., $P_0$) is the total dissipated power from the dipole with (resp., without) plasmonic structure. Dipole emission power in vacuum reads, $P_0 = \frac{|\mathbf{\mu}|^{2}}{4 \pi \varepsilon_{0} } \frac{ \omega^{4}}{3 c^{3}}$. The enhancement of the radiative LDoS is $F = P/P_0 \cdot \eta_{\rm ant}\eta_{\rm coll}$, with the antenna $\eta_{ant}$ and collection $\eta_{coll}$ efficiencies. We retrieved the emitted power $P_{\rm rad}=P\cdot \eta_{\rm ant}\eta_{\rm coll}=P\cdot \eta_{\rm rad}$ from the NPoS by a near-to-far-field transformation.\cite{yangNeartoFarFieldTransformations2016} The power collected in a solid angle of $NA = 0.9$ was acquired by a surface integral of the Poynting vector in the far field.

\textbf{QNM analysis.} Quasinormal mode analysis {based on MAN package} \cite{yan2018rigorous,wu2023modal} was performed by combining electromagnetic and \textit{weak form PDE} modules in COMSOL Multiphysics. An auxiliary field of polarization is introduced to eliminate the hidden nonlinearity in the eigenvalue problem due to the dispersive material. 

\section*{Appendix on mode overlap and conversion efficiency}
We can estimate the conversion efficiency, $\eta_{\rm conv}$, from eq. (5) in Ref.~\citenum{han2021microwave}, which refers to zero-stage (direct) conversion schemes. In this approach, the molecular vibration is modelled by its hyperpolarizability tensor, which results in a $\chi^{(2)}$ nonlinearity. 
In the limit of small cooperativities (relevant for current experiments with molecules in plasmonic nanocavities), and considering that each frequency involved is exactly resonant with a distinct cavity mode, the conversion efficiency reduces to: 

\begin{equation}
    \eta_{\rm conv}=4\,\eta^{\rm rad}_{\rm IR}\,\eta^{\rm rad}_{\rm SFG}\, \mathcal{C} \quad \text{where}  \quad  \mathcal{C}=4\, \bar{n}_{\rm VIS}\, g_{\rm 0}^2/(\kappa_{\rm IR}\kappa_{\rm SF}) \label{eq:direct_conv}
\end{equation}
is the cooperativity, with $\bar{n}_{\rm VIS}\propto P_{\rm VIS}$ the cavity mode occupancy due to the VIS pump laser. For each mode $m\in\{{\rm IR, VIS, SF}\}$ (where SF stands for sum-frequency), $\kappa_m$ is the total decay rate and $\eta^{\rm rad}_m=\kappa_m^{\rm rad}/\kappa_m$ is the coupling efficiency (the input-output formalism for plasmonics is reviewed in Ref.~\citenum{roelli2024nanocavities}). 

The vacuum coupling rate $g_{\rm 0}$ can be computed from the $\chi^{(2)}$ tensor of the medium inside the nanocavity. The full expression of the vibrational contribution to $\chi^{(2)}$ in terms of IR dipole $\mu_i$ and optical polarizability $\alpha_{jk}$ of the molecule (which enter the hyperpolarizability $\beta^{\rm (2,vib)}_{ijk}$) can be found, e.g., in Ref.~\citenum{morita2008recent}. We consider a single vibrational mode of frequency $\omega_{\rm v}$, decay rate $\Gamma_{\rm v}$, reduced mass $m_{\rm v}$ and normal coordinate $Q_{\rm v}$. We further simplify the problem by assuming that all molecules have the same orientation; then, in a cartesian coordinate system $x,y,z$, we have
\begin{equation}
    \chi^{(2)}_{ijk}(\omega_{\rm IR}) \approx \beta^{\rm (2,vib)}_{ijk}(\omega_{\rm IR}) \approx \frac{1}{2 m_{\rm v} \omega_{\rm v}}\left(\frac{\partial\alpha_{i,j}}{\partial Q_{\rm v}}\right)\left(\frac{\partial\mu_{k}}{\partial Q_{\rm v}}\right)\frac{1}{\omega_{\rm IR}-\omega_{\rm v}+i\Gamma_{\rm v}}  \label{eq:chi2_vib}
\end{equation}
where $i,j,k$ can each take any possible values among $x,y,z$. In general, a change of coordinate systems is needed to pass from the natural basis for the molecule to that of the plasmonic cavity. 
Assuming that we know the mode profiles $\bm{e}^{\rm IR}(\bm r)$, $\bm{e}^{\rm VIS}(\bm r)$, $\bm{e}^{\rm SF}(\bm r)$ at all three frequencies involved $\omega_{\rm IR}$, $\omega_{\rm VIS}$, $\omega_{\rm SF}$, then the vacuum coupling rate becomes (see, e.g., Ref.~\citenum{mckenna2020cryogenic})

\begin{equation}
 \hbar g_{\rm 0}= 2\varepsilon_0\int dV\, \chi^{(2)}_{ijk}(\bm r) {e}^{\rm IR}_{i}(\bm r) {e}^{\rm VIS}_{j}(\bm r) {e}^{\rm SF}_{k}(\bm r) A_{\rm IR}A_{\rm VIS}A_{\rm SF} \label{eq:g0dir}
\end{equation}
where the summation over repeated indices is implicit (Einstein notation) and the vacuum fluctuation amplitude $A_m$ for each mode $m$ is given by
\begin{equation}
    A_m=\sqrt{\frac{\hbar \omega_m}{2\varepsilon_0 \int \varepsilon_r(\omega_m) |\bm e^m|^2 dV}}=\sqrt{\frac{\hbar \omega_m}{2\varepsilon_0 V_m\,\max\left(\varepsilon_r|\bm e^m|^2 \right)}}
\end{equation}
where the $m$ mode volume $V_m$ is defined as
\begin{equation}
    V_m=\frac{\int \varepsilon_r(\omega_m) |\bm e^m|^2 dV}{\max\left(\varepsilon_r|\bm e^m|^2 \right)}
\end{equation}
(for simplicity, we consider an isotropic relative permittivity $\varepsilon_r$) 

In a plasmonic nanocavity, we may assume that, at the location of the molecular layer, all fields are along the same direction $\bm{u}_z$, perpendicular to the metal surface. If we express the hyperpolarizability of the molecule in the same basis, then from eq.~(\ref{eq:g0dir}) only the component $\chi^{(2)}_{zzz}$ contributes to the nonlinear interaction and we can rewrite
\begin{equation}
g_{\rm 0}=\chi^{(2)}_{zzz}\sqrt{\frac{\hbar\,\omega_{\rm IR}\,\omega_{\rm VIS}\,\omega_{\rm SF}}{2\varepsilon_0}}\, \eta_{\rm overlap}
\end{equation}
with 
\begin{equation}
    \eta_{\rm overlap}=\frac{\int_{\rm molecules} {e}^{\rm IR}_{z}(\bm r) {e}^{\rm VIS}_{z}(\bm r) {e}^{\rm SF}_{z}(\bm r) dV}{\left(V_{\rm IR}\,V_{\rm VIS}\,V_{\rm SF}\right)^{1/2}\max\left(\varepsilon_{r,\rm IR}^{1/2}|\bm e^{\rm IR}|\right){\rm max}\left(\, \varepsilon_{r,\rm VIS}^{1/2}|\bm e^{\rm VIS}|\right){\rm max}\left(\,\varepsilon_{r,\rm SF}^{1/2}|\bm e^{\rm SF}| \right) } \label{eq:overlap3}
\end{equation}
As a side note, if the $\varepsilon_r$'s can be factored out, the coupling rate reduces to 
\begin{equation}
g_{\rm 0}=\chi^{(2)}_{zzz}\sqrt{\frac{\hbar\,\omega_{\rm IR}\,\omega_{\rm VIS}\,\omega_{\rm SF}}{2\varepsilon_0\varepsilon_{r,\rm IR}\varepsilon_{r,\rm VIS}\varepsilon_{r,\rm SF}}}\, \left(V_{\rm IR}\,V_{\rm VIS}\,V_{\rm SF}\right)^{-1/2} \int_{\rm molecules}{\psi}^{\rm IR}_{z}(\bm r) {\psi}^{\rm VIS}_{z}(\bm r) {\psi}^{\rm SF}_{z}(\bm r) dV
\end{equation}
where the $\psi$'s are mode field distributions normalized to their maximal norms. Using any of the previous two equations, we see that $1/\eta_{\rm overlap}^2$ has the dimension of a volume, so we can call this quantity an effective nonlinear mode volume $V_{\rm NL}$. The smaller it is, the larger the cooperativity, and thus the larger the efficiency. 
 
 The computation of $\eta_{\rm overlap}$ or $V_{\rm NL}$ is a sufficient input for a relative measure of the conversion efficiency, independently of the nonlinear medium. We also need to estimate $\eta^{\rm rad}_{\rm IR}$ and $\eta^{\rm rad}_{\rm SF}$, and to know how efficiently the VIS pump couples into the nanocavity: 
\begin{equation}
    \frac{\bar{n}_{\rm VIS}}{P_{\rm VIS}}=\frac{\kappa_{\rm VIS}^{\rm rad}}{\hbar\omega_{\rm VIS}(\Delta_{\rm VIS}^2+\kappa_{\rm VIS}^2/4)}\propto \eta_{\rm VIS}^{\rm rad}\frac{1}{\kappa_{\rm VIS}\left(\frac{1}{4}+\frac{\Delta_{\rm VIS}^2}{\kappa_{\rm VIS}^2}\right)}
\end{equation}
where $\Delta_{\rm VIS}=\omega_{\rm pump}-\omega_{\rm VIS}$ is the detuning between the VIS pump and the nearest cavity mode, assuming a Lorentzian resonance. 

When $\omega_{\rm IR}\ll \omega_{\rm VIS}$, it is a good approximation to consider that a single QNM is mediating the interaction with both VIS and SF fields, i.e., we can perform the substitution $\bm{e}^{\rm VIS}(\bm r) \longrightarrow \bm{e}^{\rm SF}(\bm r)$ in eq.~(\ref{eq:overlap3}) and obtain:
\begin{equation}
    \eta_{\rm overlap}=\frac{\int \,{e}^{\rm IR}_{z}(\bm r) \left({e}^{\rm SF}_{z}(\bm r)\right)^2 dV}{\sqrt{\int\varepsilon_{r,\rm IR}|\bm{e}^{\rm IR}|^2 dV} \cdot{\int\varepsilon_{r,\rm SF}|\bm{e}^{\rm SF}|^2  dV}} \label{eq:overlap2}
\end{equation}

\begin{acknowledgement}

H.H. acknowledges support from the National Natural Science Foundation of China (Grant No.12204362), W.C, Z.H. acknowledge the support from the National Key Research and Development Program of China (Grant No. 2024YFA1409900) and the National Natural Science Foundation of China (Grant No. 62475071 and 52488301), C.G. acknowledges the support from the Swiss National Science Foundation (Grant No. 214993). 

\end{acknowledgement}

\section*{Author Contributions}

H.H. and Z.H. contributed equally to this work. The authors declare no competing financial interest.






\bibliography{NPiS_20241230}

\end{document}